\documentstyle[11pt,psfig]{article}
\begin{document}
\title{\huge{Prospects for the extraction of ${\mid}V_{cb}{\mid}$ from
$B_c\ {\rightarrow}\ J/\psi\ {\mu}^+\ {\nu}$ decays at LHC }
\thanks{Research partially supported by CICYT under grant AEN 93-0234}}
\author{{\bf M. Gald\'on\thanks{galdon@evalvx.ific.uv.es} $\ $ and M.A.
Sanchis-Lozano\thanks{mas@evalvx.ific.uv.es}}
\\
\it Instituto de F\'{\i}sica
 Corpuscular (IFIC) Centro Mixto Universidad de Valencia-CSIC \\
\it and \\
\it Departamento de F\'{\i}sica Te\'orica \\
\it Dr. Moliner 50, E-46100 Burjassot, Valencia (Spain) }
\maketitle
\abstract{We discuss the possibility of extracting ${\mid}V_{cb}{\mid}$
from the future measurement at LHC of the
$B_c\ {\rightarrow}\ J/\psi\ {\mu}^+\ {\nu}$ decay rate followed by the
leptonic decay  $J/\psi\ {\rightarrow}\ \mu^+\mu^-$.
We examine in detail the influence of the measurement of the muons' momenta
and vertex reconstruction in the final uncertainty of ${\mid}V_{cb}{\mid}$, as
well as the required theoretical inputs at both the production and weak decay
level, and their reliability in this regard. We stress the relevance of
a rigorous description of the fragmentation contributions to inclusive
production at large transverse momentum of heavy quarkonia in $pp$
collisions.}
\vspace{-13cm}
\large{
\begin{flushright}
  IFIC/95-31\\
  FTUV/95-29\\
  \today
\end{flushright} }
\vspace{11.5cm}
\section{Introduction}
The interplay between theory and experiment, inherent to all branches of
science, becomes crucial in high-energy physics at present. Indeed, despite
any (always welcome) unexpected discovery, the required investment of money
and time demands well-defined aims and physical objectives from the theoretical
side long time before the start of real data-taking. This is particularly
apparent at LHC experiments, in one of which (ATLAS) one of us (M.A.S.L.) is
currently involved. On the other
hand, it is a task of theorists to survey all those theoretical topics
compatible with the ultimate goals of an experiment, keeping connection with
a $\lq\lq$realistic" experimental point of view. This is, in fact, one of
the underlying motivations of the present \vspace{0.1in} work.
\par
In particular, it has been recently pointed out the feasibility of the
observation of $B_c$ mesons in the ATLAS experiment \cite{albiol}. With
respect to their production rate, setting the $b\overline{b}$ cross section
equal to 500 $\mu$b and assuming the fragmentation probability
of a $b$ quark into a $B_c$ meson
of the order of $10^{-3}$ \cite{cheung}, the expected number of $B_c$ mesons
at $\lq\lq$low" luminosity \footnote{It is foreseen that during the first
years, LHC will run at ${\cal L}\ {\approx}\ 1{\times}10^{33}
$ cm$^{-2}$ s$^{-1}$. Later on, the luminosity will increase}
is about $10^{10}$ per year ($10^7$ s) corresponding to an integrated
luminosity of \vspace{0.1in} $10$ fb$^{-1}$.
\par
In this paper we shall argue that in LHC experiments the decay
\vspace{0.06in} channel
\begin{equation}
B_c\ {\rightarrow}\ J/\psi\ {\mu}^+\ {\nu}_{\mu}
\end{equation}
followed by the leptonic decay of the $J/\psi$ into a pair of oppositely
charged muons, could provide a reliable extraction of
the mixing matrix element ${\mid}V_{cb}{\mid}$. The expected semileptonic
branching ratio ${\simeq}\ 2\%$ for channel (1) \cite{lusi} together with the
leptonic branching ratio for the $J/\psi$ into two muons
${\simeq}\ 6\%$ \cite{pdg}, yields an overall branching fraction of
order \vspace{0.1in} $10^{-3}$.\par
We next combine the last $BF$ with the acceptance \footnote{Hereafter, we
mainly
rely on technical details contained in the Technical Proposal of the ATLAS
Collaboration \cite{tp}. Those aspects directly related to the generation
of the $B_c$ signal and its background, the trigger and reconstruction
efficiencies are left to a separate publication \cite{nota}. Let
us only mention that first-level trigger muons are required to have
a transverse momentum higher than $6$ GeV/c and pseudorapidity
${\mid}{\eta}{\mid}$ less than $2.2$. The other two muons are required to
pass the cuts $p_{\bot}>3$ GeV/c and ${\mid}{\eta}{\mid}<2.5$} for the muon
trigger ${\simeq}\ 15\%$ (where the three possible triggering muons per
decay have been taken into account.) Besides, the detection of the two
remaining particles (i.e. the non-triggering muons satisfying some $p_{\bot}$
and ${\eta}$ cuts \cite{nota}) amounts to an acceptance for signal events of
${\simeq}\ 12\%$. Finally, we assume an identification efficiency for each
muon of $0.8$ \cite{tp} yielding a combined value of \vspace{0.1in}
$0.8^3\ {\simeq}\ 0.5$.\par
Thereby, the expected number of useful $B_c$ decays \footnote{Semi-electronic
$B_c{\rightarrow}J/\psi e^+ \nu$ decays, followed by the $J/\psi$ decay
into ${\mu}^+{\mu}^-$ can be considered as well. In addition, if
the muon from the semi-muonic decay (1) gives the trigger, the
event can be selected with the mode $J/{\psi}{\rightarrow}\ e^+e^-$. The
trigger acceptance for the former is ${\simeq}\ 12\%$ whereas for the latter
is ${\simeq}\ 3\%$ \cite{nota}}
could reach several $10^5$ per year at low \vspace{0.1in} luminosity.\par
Such large statistics together with the foreseen precision in the muon momentum
measurement should permit the experimental access to the kinematic region near
zero-recoil of charmonium (with respect to the $B_c$ meson) with a good
energy/momentum resolution. Moreover, very stringent cuts can be put
on events and hence, expectedly, background could be almost entirely removed
from the event sample permitting a very precise determination of the
$B_c$ lifetime. We explore all these issues in the \vspace{0.1in}
following section.

\section{Extraction of ${\mid}V_{cb}{\mid}$}
In recent times, an attempt to derive the value of the
Cabibbo-Kobayashi-Maskawa
matrix element ${\mid}V_{cb}{\mid}$ from
$\overline{B}{\rightarrow}D^{\ast}\ell\overline{\nu}$ in a (more or less)
model-independent way has been proposed \cite{neu1} within the
framework of the heavy quark effective theory (HQET)\cite{neu2}. This method
basically relies on the existence of a universal form factor, the so-called
Isgur-Wise function $\xi(w)$, depending on $w=v_1{\cdot}v_2$, variable
representing the Lorentz factor of the final hadron in the initial hadron's
frame. (Hereafter subindices 1, 2 will denote initial, final hadronic
\vspace{0.06in} quantities.)
\par
Near zero \vspace{0.06in} recoil
\par
\begin{equation}
{\lim}_{w{\rightarrow}1}\frac{1}{\sqrt{w^2-1}}\
\frac{d{\Gamma}(\overline{B}{\rightarrow}D^{\ast}\ {\ell}\overline{\nu})}{dw}\
=\ \frac{G_F^2}{4{\pi}^3}\ (m_1-m_2)^2\ m_2^3\ {\eta}_A^2\ \hat{\xi}^2(1)\
{\mid}V_{cb}{\mid}^2
\end{equation}
where ${\eta}_A$ denotes a perturbative coefficient \cite{neu2}. On the other
hand, the $\hat{\xi}(w)$ form factor absorbs those corrections arising from
inverse powers of the heavy quark masses, thus removing its universal
character. In the infinite quark mass limit, $\hat{\xi}$ coincides with the
Isgur-Wise \vspace{0.1in} function.
\par

Experimentally, one can obtain the product
$\hat{\xi}(w){\cdot}{\mid}V_{cb}{\mid}$ from the measured
differential rate $(1/{\tau}_B)dBr/dq^2$, where $q=p_1-p_2$ is the
four-momentum transfer to the leptonic system (see figure 1). As developed
by Neubert \cite{neu1}, the zero recoil point is especially
suitable for the extraction of ${\mid}V_{cb}{\mid}$ in
$\overline{B}{\rightarrow}D^{\ast}{\ell}\overline{\nu}$ decays. This is because
heavy-quark flavour symmetry and Luke's theorem determine
the normalization of $\hat{\xi}(1)$ up to second-order power corrections:
$\hat{\xi}(1)=1+{\delta}_{1/m^2}$ where the last coefficient stands for
the long-distance correction \cite{neu2}. The theoretical control of these
uncertainties permits a reliable determination of ${\mid}V_{cb}{\mid}$ at this
kinematic point where, however, there are no data due to vanishing phase
space. An extrapolation to this endpoint is thus required and the final
precision depends both on the experimental accuracy, background removal
and theoretical corrections as \vspace{0.1in} well.
\par
In order to apply a similar approach to the semileptonic channel (1) of
$B_c$ mesons we will characterize the kinematic state of hadrons by means
of their four-velocity instead of four-momenta as well. This will make sense
in our later theoretical approach as we are dealing with weakly bound
states where the constituent heavy quarks are not too far offshell, as
pointed out elsewhere \cite{masl} \vspace{0.1in} \cite{mannel}.\par
For clarity of presentation let us make below some remarks
to be developed one by one in the \vspace{0.1in} following:
\begin{itemize}
\item[{\bf a)}] The accessible $w$ range in $B_c{\rightarrow}J/\psi \mu^+\nu$
decays (see figure 2) is limited to the interval $(1,1.26)$, whereas in
$\overline{B}{\rightarrow}D^{\ast}\ell\overline{\nu}$ decays the range is
significantly wider $(1,1.5)$. The full kinematic reconstruction of the decay
is required in order to measure $w$ for each sampled \vspace{0.1in} event.
\item[{\bf b)}] The precision in the measurement of the muon's momentum in
$B$-physics at LHC experiments is of order $2\%$ \cite{tp} or
less \cite{cms}, representing an uncertainty of about $2\%$ or less in the
reconstructed momentum of the \vspace{0.1in} $J/\psi$.
\item[{\bf c)}] Background from processes like
$B{\rightarrow}J/\psi K^{\pm}$ must be taken into \vspace{0.1in} account.
\item[{\bf d)}] Contamination from $B_c{\rightarrow}X{\mu}^+\nu$ decays,
where $X={\psi}(2S),{\chi}_{c}(1P)$ stands for a higher charmonium
resonance, will be \vspace{0.1in} addressed as well.
\item[{\bf e)}] We shall consider combinatorial background coming from
$b\overline{b}$ pairs subsequently decaying into \vspace{0.1in} muons.
\item[{\bf f)}] A rigorous fragmentation framework recently developed by
Bodwin, Braaten and Lepage \cite{bodwin2} could provide a reliable evaluation
of the (high-$p_{\bot}$) production cross section of heavy quarkonia in $pp$
collisions at LHC \vspace{0.1in} energy.
\item[{\bf g)}] Precise predictions for the decay rate in the vicinity of
zero-recoil could be obtained in the framework of lattice QCD
\cite{niev} or even potential models \cite{chang} of \vspace{0.2in} hadrons.
\end{itemize}
\par
\subsection{Signal}
In this section we develop items a) \vspace{0.1in} and b):\par

{\bf a)} In order to get the $w$ differential distribution of the $B_c$
semileptonic decays, their complete kinematic reconstruction (event by event)
is required. This can be achieved despite the unknown energy of the initial
$(B_c)$ particle (whose mass is assumed to have been previously measured from
the $B_c{\rightarrow}J/\psi \pi$ decay at the LHC \cite{albiol}, or even
earlier at the Fermilab Tevatron) and the unknown momentum of
the final $\nu$, if the initial direction of the former can be experimentally
determined. Indeed, this should be the case by means of the vertexing
capability of high performance inner detectors at LHC \vspace{0.1in}
experiments.\par
In fact, a precise secondary vertex resolution for B physics has been
established as an experimental goal in the ATLAS experiment \cite{tp}
\footnote{It is foreseen the installation of an additional precision
layer at low radius during the low luminosity running}. This fact together
with the accurate reconstruction of the primary vertex at low luminosity
should permit the measurement of the $B_c$'s flight-path \vspace{0.1in}.\par
Let us write $w$ \vspace{0.06in} as
\par
\begin{equation}
w\ =\ {\gamma}_1{\gamma}_2\ (1-{\mid}\vec{v}_1{\mid}\ {\mid}\vec{v}_2{\mid}
\vspace{0.06in} \cos{\theta})
\end{equation}
where ${\gamma}_i=(1-{\mid}\vec{v}_i{\mid}^2)^{-1/2}$,
${\mid}\vec{v}_i{\mid}$ stands for the modulus of the three-velocity
of particle-$i$ and $\theta$ denotes the angle between the $J/\psi$ and
$B_c$ (reconstructed) trajectories. The influence of the geometric uncertainty
on $w$ through the cosine in the above quantity near zero-recoil
 (${\gamma}_1\ {\simeq}\ {\gamma}_2={\gamma}$) can be inferred from the
following \vspace{0.06in} expression
\par
\begin{equation}
{\delta}w\ {\simeq}\ {\mid}\vec{v}{\mid}\ {\gamma}\ \sqrt{2\ (w-1)}\
{\delta}{\theta}\ =\ \sqrt{2\ (w-1)}\ \vspace{0.06in} {\delta}{\theta}_0
\end{equation}
where $\delta{\theta}={\sigma}/(n+1){\gamma}{\mid}\vec{v}{\mid}{\tau}$ and
$\delta{\theta}_0={\sigma}/(n+1)c{\tau}$. We shall take
${\sigma}\ {\approx}\ 100\ {\mu}$m as a reasonable and safe estimate
for the combined primary+secondary vertex resolution. (The vector joining
the production and decay vertices in space could be measured with a precision
of (of the order of) $100\ \mu$m.) The $B_c$ average decay
length has been expressed in (4) as $n+1$ $\lq\lq$lifetime-units" where $n$
denotes the cut on the proper decay length, i.e. we consider only decays with
proper times larger than $n\tau$. We shall set ${\tau}\ {\approx}\ 0.75$ ps
as the expected $B_c$'s lifetime. This is halfway between
potential model predictions ($0.5$ ps) \cite{lusi}
and QCD sum rules expectations ($1$ ps) \vspace{0.1in} \cite{paver}.
\par
It is quite remarkable that the accuracy of $w$ becomes independent
of kinematic factors since a more precise measurement of the angle due to a
longer decay length because a larger Lorentz factor,
is compensated by a greater influence on ${\delta}w$.
On the other hand, notice that the uncertainty ${\delta}w$ increases
with $w$. This behaviour may be qualitatively understood since a lower
recoil implies a smaller $\theta$ angle (${\theta}=0$ in the limit $w=1$), thus
a smaller uncertainty through \vspace{0.1in} $\cos{\theta}$.
\par
Tentatively we shall set $n=7$ (thus ${\delta}{\theta}_0\ {\simeq}\ 56$ mrad)
amounting to a reduction factor $10^{-3}$ of the original sample of
triggered and reconstructed events.
In fact, this choice represents a compromise: the requirement of longer $B_c$
flight-paths by imposing higher $n$ values should certainly diminish the
experimental uncertainty on $\theta$ but would imply a further loss of
statistics, thereby increasing statistical \vspace{0.2in} errors.
\par
{\bf b)} With regard to the influence on $w$ from the muon momentum resolution
we \vspace{0.06in} obtain
\par
\begin{equation}
{\delta}w\ {\simeq}\ 2\ (w-1)\ \frac{\delta\gamma}{\gamma}
\end{equation}
Setting ${\delta}{\gamma}/{\gamma}\ {\simeq}\ 2\%$, we notice that
numerically its actual influence on $w$ is smaller than that due to the
uncertainty through the angle $\theta$.
\subsection{Background}
We address below those aspects related to contamination of the signal (1)
by either $B$ decays or combinatorial \vspace{0.2in} background:\par

{\bf c)} Background from decays like $B{\rightarrow}J/\psi K^{\pm}$
may fake the $B_c$ signal if the $K$ is misidentified as a
muon. This could happen due to punchthrough, for example, if the kaon decays
to a muon in flight before it reaches the calorimeter. Let us remark, however,
that such muon could be rejected from the stand-alone measurement of its
momentum in the air-core toroid muon detector of ATLAS, once compared to the
(larger) momentum of the parent kaon independently measured in the inner
\vspace{0.1in} detector.\par
In addition, there are some possible strategies to remove this
kind of contamination by firstly requiring {\em coplanarity} among
the direction of the parent meson, the reconstructed trajectory of the
$J/\psi$ and the $\mu$ track. In fact, this requirement amounts to
no missing transverse momentum with respect to the direction of the
decaying particle. Besides, the invariant mass of the final particles
would be compatible with the $B$ decay \vspace{0.1in} hypothesis.\par
There is, however, another potential source of contamination from
$B{\rightarrow}J/\psi K^{\ast}$ decays occurring again at a comparable rate
as the signal. The $K^{\ast}$ subsequent decay into $K^{\pm}$ and an undetected
${\pi}^0$ would make useless the last invariant mass condition. Nevertheless,
since the momentum of the $K^{\pm}$ in the $K^{\ast}$ rest frame is 291 MeV
\cite{pdg}, a lower cut of about $300$ MeV on the component of the $K^{\pm}$
momentum perpendicular to the plane defined by the decaying particle and
the $J/\psi$ should remove all that background with a good acceptance
($\simeq\ 80\%$) for the signal \cite{nota}. In fact, this cut is compatible
with the precision on the transverse momentum due to the flight-path
reconstruction. Likewise, other sources of background from channels like
$B{\rightarrow}X_cK^{\pm}$ where $X_c$ denotes charmonium resonances
subsequently decaying into $J/\psi$ and neutrals, should be rejected as well
by means of this \vspace{0.2in} cut.
\par
{\bf d)} Contamination coming from $B_c{\rightarrow}X{\mu}^+\nu$ decays where
$X={\psi}(2S),{\chi}_c(1P)$, followed by the resonance radiative decay into
a $J/\psi$, would contaminate the signal (1). Indeed, one cannot get rid of
this contamination by imposing to the semileptonic decays a missing mass
constraint, contrary to
$\overline{B}{\rightarrow}D^{\ast}{\ell}\overline{\nu}$
in a $e^+e^-$ collider. However, it may be possible to detect
the photon in the electromagnetic calorimeter, with typical energy of
few GeV, predominantly emitted within a
certain cone along the direction of the $J/\psi$. In this way those events
yielding an invariant mass compatible with any of the above-mentioned
resonances
would be removed from the event sample. Nevertheless, before any serious
simulation analysis about this possibility had been performed, we prefer to
stay at more conservative grounds and consider provisionally
the emitted $\gamma$ as useless. Therefore we must
evaluate the degree of contamination of the signal \vspace{0.1in} sample.\par
We tentatively adopt experimental data on $B$ decays by making the reasonable
assumption that the $B_c$ should decay semileptonically into higher resonances
and $J/\psi$ with a relative ratio of order $2.7/7$ \cite{pdg}. Further,
using crudely an $\lq\lq$average" $BF\ {\simeq}\ 0.16$ for the decay of
the charmonium resonances into $J/\psi$ and neutrals \cite{pdg}, we conclude
that the expected contamination of the signal from all the above channels
represents about $6\%$ hence below the statistical
fluctuations of the sampled signal events as we shall \vspace{0.1in} see.
\par
{\bf e)} Combinatorial background is generated by the simultaneous coexistence
of three muons in events. For instance this could happen if a $J/\psi$ is
produced from a $B$ whereas the other muon comes from a different $b$ quark
(thus from two spatially well separated vertices). This kind of contamination
should be drastically suppressed by means of the seven-lifetimes cut imposed
to the reconstructed secondary vertices. Furthermore, prompt
$J/\psi$'s produced at the primary $pp$ interaction should be
practically entirely removed as a background source by requiring a combined
cut on the ${\chi}^2$ of the fitted secondary vertex and the decay length
\cite{albiol} \cite{nota}.

\subsection{Theory inputs and determination of ${\mid}V_{cb}{\mid}$}
Next we examine in detail the possibility of extracting ${\mid}V_{cb}{\mid}$
from experimental \vspace{0.1in} data.\par
{\bf f)} Let us start by recalling \vspace{0.1in} that
\begin{equation}
\frac{d{\Gamma}(B_c{\rightarrow}J/\psi\ {\mu}^+\nu)}{dw}\
=\ \frac{1}{\tau}\
\frac{dBr(B_c{\rightarrow}J/\psi\ {\mu}^+\nu)}{dw}
\end{equation}
\vspace{0.06in}
\par
The differential branching ratio $dBr/dw$ can be written
\vspace{0.06in} as
\begin{equation}
\frac{dBr(B_c{\rightarrow}J/\psi\ {\mu}^+\nu)}{dw}\ =\
\frac{1}{N(B_c)}\frac{dN(B_c{\rightarrow}J/\psi\ {\mu}^+\nu)}{dw}
\end{equation}
where $dN/dw$ represents the number of semileptonic decays per unit
of $w$ for a certain integrated luminosity, once corrected by detection
efficiency; $N(B_c)$ is the corresponding total yield
of $B_c$'s produced for such integrated luminosity. (Notice that $\tau$ could
be determined accurately from the same collected sample of semileptonic
$B_c$ \vspace{0.1in} decays.)
\par
Let us observe that the exclusive decay mode $B_c{\rightarrow}J/\psi \pi$
to be detected at LHC \cite{albiol} cannot be employed for normalization
purposes as far as its $BF$ would  not be (obviously yet)
experimentally known. On the other hand, $B_c$ decay modes
involving the $c$-quark (which in principle would permit the
measurement of ratios of CKM matrix elements in combination
with $b$ decays) will be hardly observable among
the huge background of a hadron \vspace{0.1in} collider.
\par
Therefore, in order to get $N(B_c)$ it will be convenient to compare the
$B_c$ production rate to the prompt ${\psi}'$ yield in $pp$ collisions,
for the same integrated luminosity. \vspace{0.06in} Therefore
\begin{equation}
N(B_c)\ =\ \frac{\sigma(pp{\rightarrow}B_c+X)}
{\sigma(pp{\rightarrow}{\psi}'+X)}\ {\times}\
\frac{N({\psi}'{\rightarrow}{\mu}^+{\mu}^-)}
{{\epsilon}_{{\psi}'}{\cdot}BF({\psi}'{\rightarrow}{\mu}^+{\mu}^-)}
\end{equation}
where $N({\psi}'{\rightarrow}\mu^+\mu^-)$ stands for the number of prompt
${\psi}'$ experimentally observed through the muonic decay mode and
${\epsilon}_{{\psi}'}$ denotes its detection efficiency. Contamination
from weak decays of $B$ mesons into ${\psi}'$ states should be efficiently
removed by means of the vertex capability of the inner
\vspace{0.1in} detector \cite{albiol} \cite{nota}.
\par
Therefore we suggest normalizing the $B_c$ sample with the aid of the
${\psi}'$ yield through expression (8). The ${\psi}'$ state is preferable to
the $J/\psi$ mainly because the former expectedly should not be fed down by
higher charmonium states \footnote{See Ref. \cite{close1} for an alternative
explanation of the observed ${\psi}'$ surplus found in Tevatron. However this
mechanism would require an uncomfortable large radiative $BF$ for higher
resonances}. Instead, data released by Tevatron shows that
the $J/\psi$ indirect production through ${\chi}_c$ intermediate states may
easily overwhelm direct production. Let us also
remark that the production of the ${\psi}'$ resonance in $p\overline{p}$
collisions is more than one order of magnitude larger than initially
expected \cite{ma} \vspace{0.1in} \cite{greco}.
\par
At sufficiently large $p_{\bot}$ \footnote{We find an average
$p_{\bot}\ {\approx}\ 20$ GeV/c for $B_c$ mesons simply passing the
kinematics cuts on $p_{\bot}$ and ${\mid}{\eta}{\mid}$ of the decay muons
\cite{nota}}, one reasonably expects that the main
contribution to heavy quarkonia production comes from the splitting of
a heavy quark or a gluon. Indeed, even though the fragmentation process
is of higher order in ${\alpha}_s$ than the $\lq\lq$conventional" leading
order diagrams \cite{shuler}, the former is enhanced by powers of
$p_{\bot}/m_Q$ relative to the latter \cite{cheung} \cite{roy}. Although the
situation is still controversial in the literature, there are strong
indications
from very detailed calculations (see \cite{leike} \cite{greco2} and references
therein) that fragmentation indeed dominates for large enough $p_{\bot}$. In
the following, we first examine some color-singlet mechanisms, expected to
contribute largely to the high-$p_{\bot}$ inclusive production of prompt heavy
\vspace{0.1in} quarkonia.
\par
According to the set of papers in \cite{cheung}, perturbative QCD can
provide a reliable calculation for the fragmentation function of a
high-$p_{\bot}$ parton into heavy quarkonia with only few input parameters:
the charm and bottom masses, and the square of the radial wave function
of heavy quarkonium at the origin ${\mid}R(0){\mid}^2$. Note that the largest
uncertainty for normalization purposes comes from the (third power of the)
charm mass in the fragmentation function (see \cite{cheung} for explicit
expressions). However, we are only interested in the \vspace{0.1in} ratio
\begin{equation}
r\ =\ \frac{\sigma(pp{\rightarrow}B_c+X)}{\sigma(pp{\rightarrow}{\psi}'+X)}
\end{equation}
where each cross section can be written as a convolution of the parton
distribution functions of the colliding protons, the cross sections for the
hard subprocesses leading to the fragmenting gluon or heavy quark and the
respective fragmentation functions \cite{cheung} \cite{hep}. Thus notice
that the fragmentation
functions $D_{\overline{b}{\rightarrow}B_c}$, $D_{g{\rightarrow}B_c}$ and
$D_{c{\rightarrow}{{\psi}'}}$, $D_{g{\rightarrow}{{\psi}'}}$ \cite{gluon},
which themselves are independent of the parton-level subprocesses, appear
combined as a ratio. Therefore, those uncertainties coming from the
common overall factor $m_c^3$ automatically
cancel each other. Moreover, those uncertainties introduced by
the parton distribution functions also should significantly diminish
in $r$ as well, especially when a cut on the transverse momentum
of heavy quarkonia is required. Still one must evaluate
\begin{equation}
{\kappa}_0\ =\ \frac{{\mid}R_{B_c}(0){\mid}^2}{{\mid}R_{{\psi}'}(0){\mid}^2}
\end{equation}
In fact there is the possibility of expressing Eq. (10) according
to the general factorization analysis of \cite{bodwin2} in a more rigorous
way \vspace{0.1in} as
\begin{equation}
{\kappa}_0\ =\ \frac{<0{\mid}O_1^{B_c}(^1S_0){\mid}0>}
{<0{\mid}O_1^{{\psi}'}(^3S_1){\mid}0>}
\end{equation}
where $O_n^X$ are local four quark operators and the matrix elements can
be evaluated from NRQCD. A similar expression holds for the
$B_c^{\ast}$. (Moreover, the denominator can be determined from
the measured leptonic width of the \vspace{0.1in} ${\psi}'$.)\par
Recently, an additional color-octet fragmentation mechanism \cite{fleming}
has been suggested in order to reconcile the experimental results on inclusive
${\psi}'$ production with theoretical predictions. This mechanism assumes
the creation from gluon fragmentation of a $c\overline{c}$ pair in a
color-octet state, in analogy to ${\chi}_c$ production \cite{ma}. Then a new
nonperturbative parameter $H'_{8({\psi}')}$ is required, appearing
in $r$ combined as the dimensionless \vspace{0.1in} factor
\begin{equation}
{\kappa}_{8({\psi}')}\ =\ \frac{2{\pi}m_c^2H'_{8({\psi}')}}
{{\mid}R_{{\psi}'}(0){\mid}^2}\ =\ \frac{<0{\mid}O_8^{{\psi}'}(^3S_1){\mid}0>}
{<0{\mid}O_1^{{\psi}'}(^3S_1){\mid}0>}
\end{equation}
where we remark that it can be expressed in terms of NRQCD matrix
\vspace{0.1in} elements \cite{bodwin2} \footnote{$H'_{8({\psi}')}$ can be
phenomenologically estimated from $B$ or $\Upsilon$ decays as well \cite{ma}
\cite{tot} \cite{bodwin}. On the other hand, according to a nonrelativistic
quark model $H'_8$ is related to a fictitious color-octet wave
function at the origin}.
\par
Furthermore, if we do not neglect the contribution to the total
$B_c$ production of those orbital excitations like $P$-wave states, new
fragmentation functions $D_{\overline{b}{\rightarrow}\overline{b}c(P)}$
come into play \cite{cheung2}. Accordingly, some new ${\kappa}_i$ factors
will appear which can be conveniently expressed, for instance in
analogy to Eqs. (11) and (12), as:
\begin{equation}
{\kappa}_{1(B_c)}\ =\ \frac{2{\pi}\hat{m}^2H_{1(\overline{b}c)}}
{{\mid}R_{B_c}(0){\mid}^2}\ {\approx}\ \frac{1}{\hat{m}^2}\
\frac{<0{\mid}O_1^{B_c}(P_J){\mid}0>}{<0{\mid}O_1^{B_c}(^1S_0){\mid}0>}
\end{equation}
and
\begin{equation}
{\kappa}_{8(B_c)}\ =\ \frac{2{\pi}\hat{m}^2H'_{8(\overline{b}c)}}
{{\mid}R_{B_c}(0){\mid}^2}\ {\approx}\ \frac{1}{\hat{m}^2}\
\frac{<0{\mid}O_8^{B_c}(^3S_1){\mid}0>}{<0{\mid}O_1^{B_c}(^1S_0){\mid}0>}
\end{equation}
where $\hat{m}$ is the reduced mass of the $b$ and $c$ quarks. The
long-distance $H_1$ parameter \cite{bodwin} is related to the square of
the derivative of the color-singlet wave function at the origin \vspace{0.1in}.
\par
Higher $B_c$ resonances like $D$-wave states might be further taken
into account, introducing new nonperturbative parameters \cite{chowise}
involving higher derivatives of the wave function at the origin, or the
NRQCD matrix element \vspace{0.1in} analogues.
\par
In sum, the evaluation of the ${\kappa}_i$ parameters would allow one to
complete the computation of the ratio $r$ and thereby to obtain
from the experimental measurement of the ${\psi}'$ yield, the total number
of $B_c$ events for an integrated luminosity with the aid of
expression \vspace{0.1in} (8).
\par
We next address the hadronic transition between the initial
$B_c$ and final $J/\psi$ \vspace{0.2in} particles.
\par
{\bf g)} From the theoretical side the hadronic transition
$B_c{\rightarrow}J/\psi$ involves two heavy-heavy systems. Although
originally HQET was only applied to heavy-light bound states, there is a
growing belief that it can be cautiously applied to the former \cite{masl}
\cite{mannel}. Indeed, spin symmetry should be still valid as a first order
approach in the dynamics of heavy-heavy hadrons. This allows the introduction
of the analogue of the Isgur-Wise function for transitions involving doubly
heavy hadrons, denoted by ${\eta}_{12}(v_1{\cdot}v_2)$. At the non-recoil point
$v_1=v_2=v$, we shall \vspace{0.06in} write \cite{masl}
\begin{equation}
<\ J/\psi{\mid}\ A^{\mu}\ {\mid}B_c\ >\ =\
2\ {\eta}_{12}(1)\ \sqrt{m_1m_2}\ {\varepsilon}_2^{\ast\mu}
\end{equation}
where $A^{\mu}=\overline{c}{\gamma}^{\mu}{\gamma}_5b$ stands for
the axial-vector current and ${\varepsilon}_2^{\mu}$ represents the
four-vector polarization of the \vspace{0.1in} $J/\psi$.\par
Following similar steps as in the $B$ decay into $D^{\ast}$, we
write the analogue of equation (2) \vspace{0.06in} as
\begin{equation}
{\lim}_{w{\rightarrow}1}\frac{1}{\sqrt{w^2-1}}\
\frac{d{\Gamma}(B_c{\rightarrow}{J/\psi}\ {\mu}^+{\nu})}{dw}\ =\
\frac{G_F^2}{4{\pi}^3}\ (m_1-m_2)^2\ m_2^3\ {\eta}_{12}^2(1)\
{\mid}V_{cb}{\mid}^2
\end{equation}
\par
It is of key importance from the theoretical side in our proposal, the
existence of a single form factor in the above expression to be determined
theoretically in a rigorous manner. Observe that this is valid only in a
neighbourhood of $w=1$, however. Deviations from the strict requirement of
the spin symmetry lead to the appearance of new form factors in the hadronic
matrix element \cite{neu2}. Even though they can be absorbed in a redefinition
of ${\eta}_{12}(w)$ this does not mean that we could ignore them if
${\mid}V_{cb}{\mid}$ has to be extracted away from zero \vspace{0.1in}
recoil.\par
Note that ${\eta}_{12}(1)$ may be interpreted in first approximation
as the overlap of the initial and final hadron wave functions \cite{masl}
\cite{mas}. Due to the intrinsic nonrelativistic nature of its heavy
constituents, nonrelativistic QCD on the lattice or even
refined potential based models of hadrons, could be applied
in order to get ${\eta}_{12}$ and its slope at zero recoil
in a reliable \vspace{0.1in} way.\par
This fact in conjunction with the presumed accuracy of the $J/\psi$
momentum measurement suggests determining ${\mid}V_{cb}{\mid}$ in a similar
way as in the method proposed by Neubert \cite{neu1} through the
\vspace{0.1in} $\overline{B}{\rightarrow} D^{\ast}\ell\overline{\nu}$ decay.
\par
Observe that from an experimental point of view the
$D^{\ast}$ kinematic reconstruction should suffer from larger uncertainties
through the cascade decays: $D^{\ast}{\rightarrow}D\ \pi$ followed by
the $D$ decay into final states with only charged particles. Conversely, the
$J/\psi$ decay into ${\mu}^+{\mu}^-$ is much cleaner by all means. Of
course, the environment is much dirtier in a hadron collider than in a $e^+e^-$
factory \footnote{One may conjecture about the possibility of producing
$B_c\overline{B}_c$ pairs in a future ${\mu}^+{\mu}^-$ collider \cite{cline}}.
Nevertheless, since the signal consists of three muons the $B_c$ decay can be
disentangled from the huge hadronic background by means of adequate kinematics
cuts and constraints \cite{nota}, at least at low \vspace{0.1in} luminosity.
\par
In order to make a crude estimate of the slope of the analogous Isgur-Wise
function ${\eta}_{12}$ at zero recoil, we shall make use of a nonrelativistic
approach. Specifically we employ the ISGW \cite{isgw} model which should give
sensible results in the case under examination.
The overlap wave function then \vspace{0.06in} reads
\begin{equation}
{\eta}_{12}(t)\ =\ \biggl[\frac{2{\beta}_1{\beta}_2}
{{\beta}_1^2+{\beta}_2^2}\biggr]^{3/2}\
exp\ \biggl[-\frac{m_c^2}{2m_1m_2}\frac{t_m-t}
{k^2({\beta}_1^2+{\beta}_2^2)}\ \biggr]
\end{equation}
where $t=q^2$, $t_m=(m_1-m_2)^2$ and ${\beta}_i$ denotes a variational
parameter
corresponding to the $i$-meson obtained by adjusting the hadronic spectra;
$k$ is a parameter introduced somewhat by hand, usually
set equal to \vspace{0.1in} 0.7.\par
Now let us introduce the slope parameter ${\rho}_{12}^2$ of
the ${\eta}_{12}(w)$ function near zero recoil \vspace{0.06in} as
\begin{equation}
{\eta}_{12}(w)\ =\ {\eta}_{12}(1)\ \{\ 1\ -\
{\rho}_{12}^2\ (w-1)\ +\ {\cal O}[(w-1)^2]\ \}
\end{equation}
\par
Since $t_m-t=2m_1m_2(w-1)$, we find that ${\rho}_{12}^2$
can be expressed according to the ISGW model \vspace{0.06in} as
\begin{equation}
{\rho}_{12}^2\ =\ \frac{m_c^2}{2{\beta}_{12}^2}
\end{equation}
where ${\beta}_{12}^2=k^2({\beta}_1^2+{\beta}_2^2)/2$. For transitions
between heavy-light mesons, ${\beta}_1={\beta}_2={\beta}$ and putting
$k=1$ one quickly recovers the result of \cite{close}:
${\rho}^2=m_q^2/2{\beta}^2$. (The relativistic correction due to the boosted
four momentum of the light quark $q$ should not affect ${\rho}_{12}^2$ so
\vspace{0.1in} much.)\par
Setting the parameters in Eq. (19) equal to some typical numerical values
for $B_c$ and $J/\psi$ states, ${\beta}_1=0.82$ GeV, ${\beta}_2=0.66$ GeV,
$m_c=1.5$ GeV and letting $k$ vary between 0.7 and 1, we estimate a
\vspace{0.06in} range
\begin{equation}
{\rho}_{12}^2\ \approx\ 2\ -\ 4 \vspace{0.06in}
\end{equation}
representing a higher value than ${\rho}^2$ (of order unity) for transitions
between $B$ and  $D^{\ast}$ mesons \cite{neu1}, as otherwise expected
\cite{shif}. (Of course, this range of ${\rho}_{12}^2$ values must be
considered as merely indicative.) Note also that one
expects the curvature of ${\eta}_{12}(w)$ at $w=1$ for transitions between
heavy-heavy hadrons to be larger than that of ${\xi}(w)$ between heavy-light
ones. This implies the necessity of a closer access to the zero recoil point
for the accuracy of the Taylor expansion of Eq. (18) as is indeed the
\vspace{0.1in} case.
\par
For the purpose of illustration, we have depicted in figure 3 some expected
typical points and error bars for the decay
$B_c{\rightarrow}J/\psi({\rightarrow}{\mu}^+{\mu}^-)\ {\mu}^+\nu$ for
a definite prediction of the ratio of production cross sections
\vspace{0.1in} $r$ in Eq. (9).\par
Under this condition, let us remark that the uncertainty on
$w$ (i.e the horizontal error bars) is of systematic nature, mainly
coming from the $B_c$ flight-path reconstruction. With regard to the vertical
coordinate ${\eta}_{12}(w){\cdot}{\mid}V_{cb}{\mid}$, we have assumed that
its uncertainty is essentially statistical, besides the contamination by
decays into higher charmonium resonances which are included
as well in the vertical error bars.

\section{Summary and discussion}
This work concerns $B_c$ mesons to be copiously produced at LHC next century.
We have argued that it should be possible to reconstruct those semileptonic
decays into a $J/\psi$ near zero recoil, followed by its consequent decay into
a muon pair. The fact that the $J/\psi$ is a vector particle should lead
to a mild falling off of the number of events in the kinematic region near
zero recoil as in $\overline{B}{\rightarrow}D^{\ast}\ell\overline{\nu}$. (This
is because of the $s$-wave contribution to these decays, absent for a final
pseudoscalar hadron as in the $\overline{B}{\rightarrow}D \ell \overline{\nu}$
\vspace{0.1in} channel.)
\par
The signature of the decay
$B_c{\rightarrow}J/\psi({\rightarrow}{\mu}^+{\mu}^-)\ {\mu}^+\nu$,
consisting of three muons coming from a common secondary vertex could be
disentangled from the huge hadronic background. The excellent accuracy in the
measurement of the muons momenta would lead to a precise determination of the
$J/\psi$ momentum and energy. In addition, a good $3D$ vertex reconstruction
capability is determinant for the full kinematic reconstruction of events. We
have shown that, in fact, it is
${\delta}{\theta}_0={\sigma}/(n+1)c\tau$ one of the parameters
that actually matters in the expected accuracy of the differential $w$
distribution of sampled \vspace{0.1in} events.\par
We also addressed some background sources, either from real $B$ decays but
faking $B_c$ ones, or from true $B_c$ semileptonic decays into
higher charmonium resonances. Several strategies to remove all
these contaminations were \vspace{0.1in} proposed.
\par
 From the theoretical point of view, the heavy quark spin symmetry permits
the introduction of the analogue of the Isgur-Wise function ${\eta}_{12}$ for
transitions between doubly heavy mesons. Deviations from the spin symmetry
imply the appearance of new form factors in the hadronic matrix element
which, however, do not contribute at zero recoil. Hence, at this kinematic
point, only a single form factor (though including itself spin breaking
effects) has to be determined for the $B_c{\rightarrow}J/\psi$ \vspace{0.1in}
transition.
\par
Experimental data would permit to obtain the product
${\eta}_{12}(w){\cdot}{\mid}V_{cb}{\mid}$ near zero recoil in a more or less
accurate way according to the accuracy of the differential decay rate
$d{\Gamma}/dw$ itself. The measurement of the latter requires
the normalization of the $B_c$ semileptonic events to
the total number of the $B_c$ sample. (In other words
the knowledge of the absolute branching fraction for the semileptonic
decay is required.) To this end, one can rely on the fragmentation
approach providing in a rigorous way the ratio of the yield of
$B_c$ mesons relative to the prompt $\psi(2S)$'s produced in $pp$ collisions
(see Eq. (8)). As far as we are interested in their relative production
rate $r$, the uncertainty should be much less dependent on
theoretical or phenomenological inputs (like heavy quark masses) than in each
production rate \vspace{0.1in} separately.
\par
We have shown that a set of long-distance dimensionless factors (denoted by
${\kappa}_{0,1,8}$) involving ratios of nonperturbative parameters like
${\mid}R(0){\mid}^2$, $H_1$ and $H'_8$ must be determined to make a reliable
prediction of $r$. Even if their evaluation from NRQCD \cite{bodwin2} is
still in its infancy, lattice techniques should permit in a not too
far future to go beyond model estimates, including relativistic
corrections, in a systematic and reliable \vspace{0.1in} way.
\par
In addition, forthcoming $e^+e^-$ factories and Tevatron running
at higher luminosity should improve considerably the accuracy of the measured
$BF$ of the $B$ and $\Upsilon$ decay modes involving $P$-states of
charmonium, improving the experimental information on such parameters.
Let us also remark that the fragmentation hypothesis can be checked with
experimental data from $pp$ or $p\overline{p}$ collisions, for
instance through the transverse momentum dependence of heavy
quarkonia, its polarization and association to jets of light
\vspace{0.1in} hadrons.\par
Once the differential rate $d{\Gamma}/dw$ is achieved to be experimentally
known, an extrapolation to
the $w=1$ endpoint, either linear or quadratic (i.e. including the curvature)
would yield ${\eta}_{12}(1){\cdot}{\mid}V_{cb}{\mid}$. Consequently, the
knowledge of ${\eta}_{12}(1)$ should provide ${\mid}V_{cb}{\mid}$. Let
us stress that this calculation and the evaluation of the fragmentation
probability into heavy quarkonium, both involving heavy quarks, rely
essentially on similar theoretical grounds and could be derived from first
principles \vspace{0.1in} \cite{bodwin2}.
\par
Of course, one may consider as well the possibility of choosing a kinematic
point different from $w=1$ to obtain ${\mid}V_{cb}{\mid}$. This requires,
however, the introduction of new $independent$ form factors (i.e. entering
in the hadronic matrix element with different Lorentz structures) whereas at
zero recoil one must evaluate all these corrections to a single form factor.
Furthermore, from an experimental point of view the non-recoil point is also
more favourable. For example, a possible choice might be the maximum recoil
point of the  $J/\psi$ at $w=1.26$ ($q^2\ {\simeq}\ 0$), where several models
predict the wave function overlap \cite{korner}. Nevertheless, notice from
figure 3 that, despite smaller vertical (statistical) error bars, horizontal
(systematic) ones are larger, hence limiting a close access
to this point and thereby making more advantageous the
extrapolation to the $w=1$ \vspace{0.1in} end-point.\par
Lastly we make some final \vspace{0.1in} comments:
\begin{itemize}
\item From the experimental viewpoint, observe that if the secondary
vertex resolution could be improved and/or the lifetime of the $B_c$ was
closer to $1$ ps than to $0.5$ ps, the experimental accuracy near zero
recoil improves considerably. Thus, we show in figure 4 possible plots for
different angular \vspace{0.1in} resolutions.
\item It is important to stress that the quantity to be
experimentally measured with a large accuracy from
$B_c{\rightarrow}J/\psi\ {\mu}^+\nu$ decays can be written as the linear
combination:
\[ r^{1/2}\ {\times}\ {\eta}_{12}(1)\ {\times} {\mid}V_{cb}{\mid} \]
\end{itemize}
\vspace{0.1in}
\par
In summary, we suggest to keep an open mind on the possibility of an
alternative determination of ${\mid}V_{cb}{\mid}$ from semileptonic decays
of $B_c$ mesons at LHC. Even if at the time LHC will start to run, B
factories would have already provided a (still) more precise value
of ${\mid}V_{cb}{\mid}$ than at present via the semileptonic
$B$ decay, an independent determination of it should bring a valuable
cross-check. On the other
hand, a lot of activity is being devoted to the analysis of inclusive
production of prompt charmonium resonances at Fermilab. Therefore we conclude
that if the fragmentation approach to describe the production of
heavy quarkonia at the large transverse momentum domain is accurately
tested (and tuned) from experimental data, $B_c$ semileptonic
decays could be competitive with the $B$ \vspace{0.1in} ones.\par
Alternatively, one can turn the question round and
consider ${\mid}V_{cb}{\mid}$ as a well-known parameter thus verifying
QCD calculations, either at the production or at the weak decay level.
Indeed, fragmentation into heavy quarkonia offers an interesting check
of perturbative QCD and a deep insight into the nonperturbative dynamics in
the hadronic formation. Besides, a precise knowledge of the $B_c$
production rate is a necessary condition for the experimental
measurement of the absolute branching fraction of any of its decay
\vspace{0.2in} modes.
\subsubsection*{Acknowledgments}
We are especially indebted to P. Eerola and N. Ellis, and the B physics group
of the ATLAS Collaboration for comments, suggestions and an
encouraging attitude. Discussions on partial aspects of this work
with V. Gim\'enez, M. Neubert,  A. Pineda, O. P\`{e}ne and J. Soto are
acknowledged as well.
\newpage

\begin{figure}[end]
\centerline{\vbox{
\psfig{figure=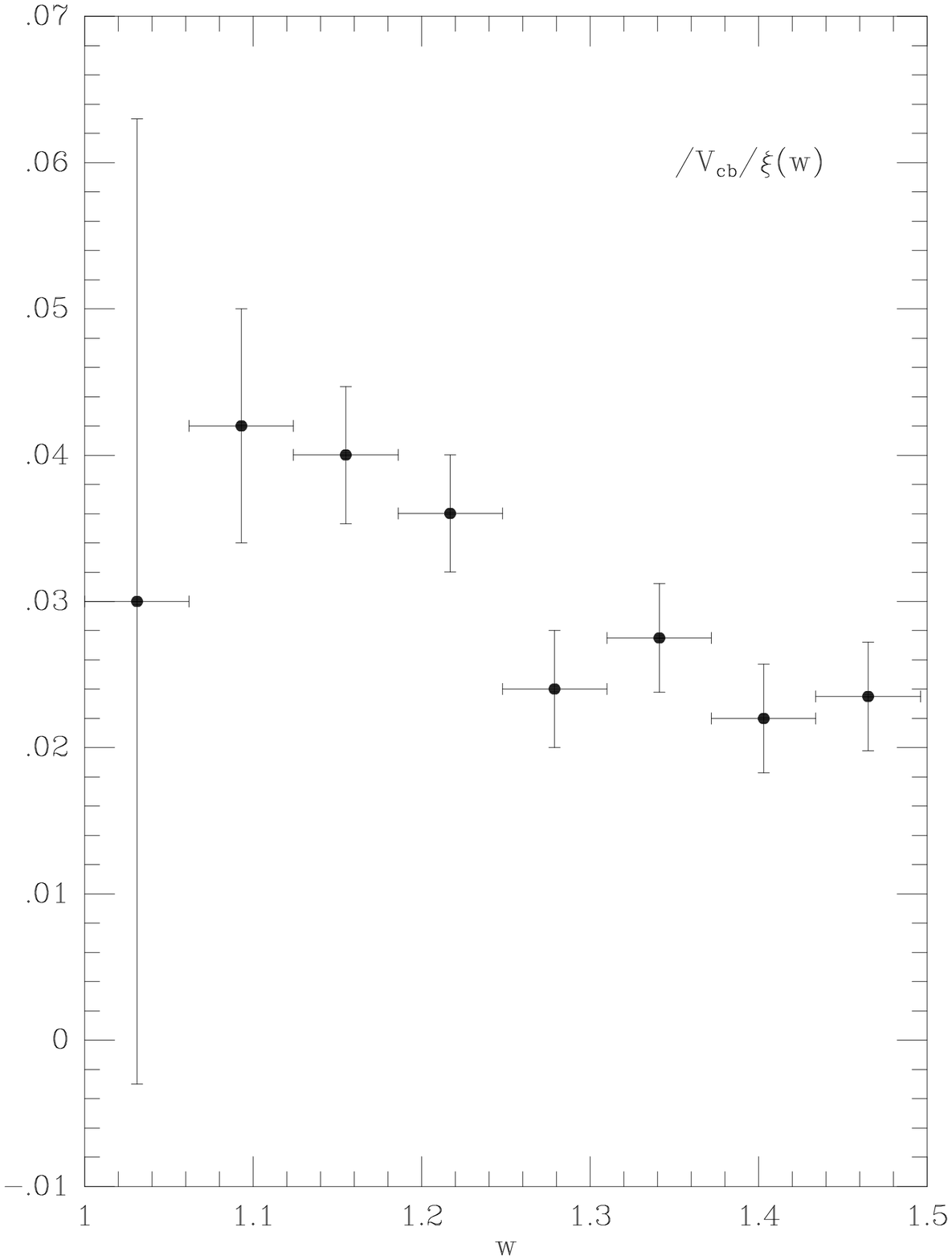,height=11cm,width=11cm}
\psfig{figure=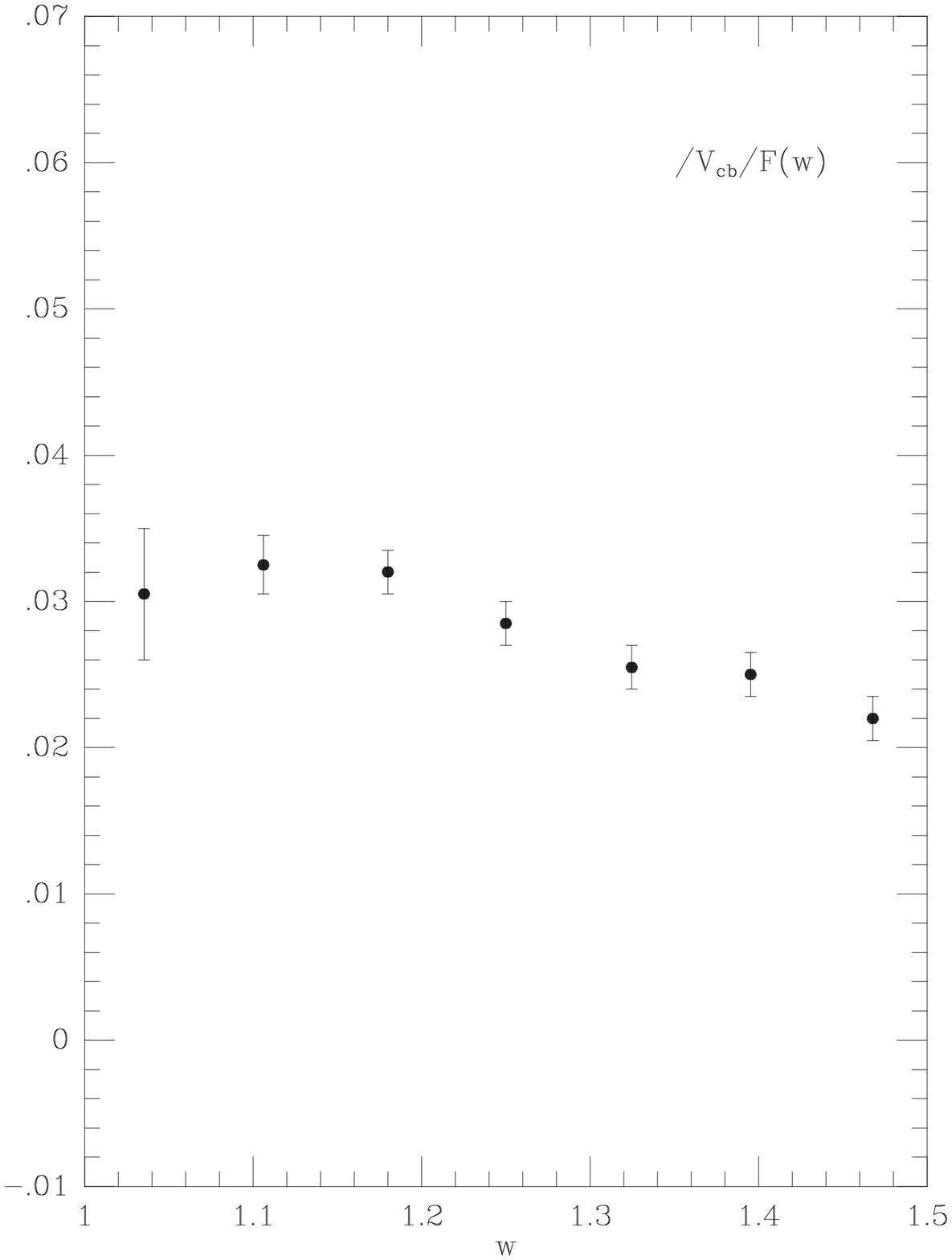,height=11cm,width=11cm}
}}
\caption[\bf Figure 1]{$\hat{\xi}(w){\cdot}{\mid}V_{cb}{\mid}$
distribution derived from the $q^2$-spectrum of the decay
$\overline{B}{\rightarrow}D^{\ast}{\ell}\overline{\nu}$ measured
by ARGUS \cite{argus} and ${\eta}_A\hat{\xi}(w){\cdot}{\mid}V_{cb}{\mid}$
measured by CLEO \cite{cleo}.}
\end{figure}

\newpage

\begin{figure}[end]
\psfig{figure=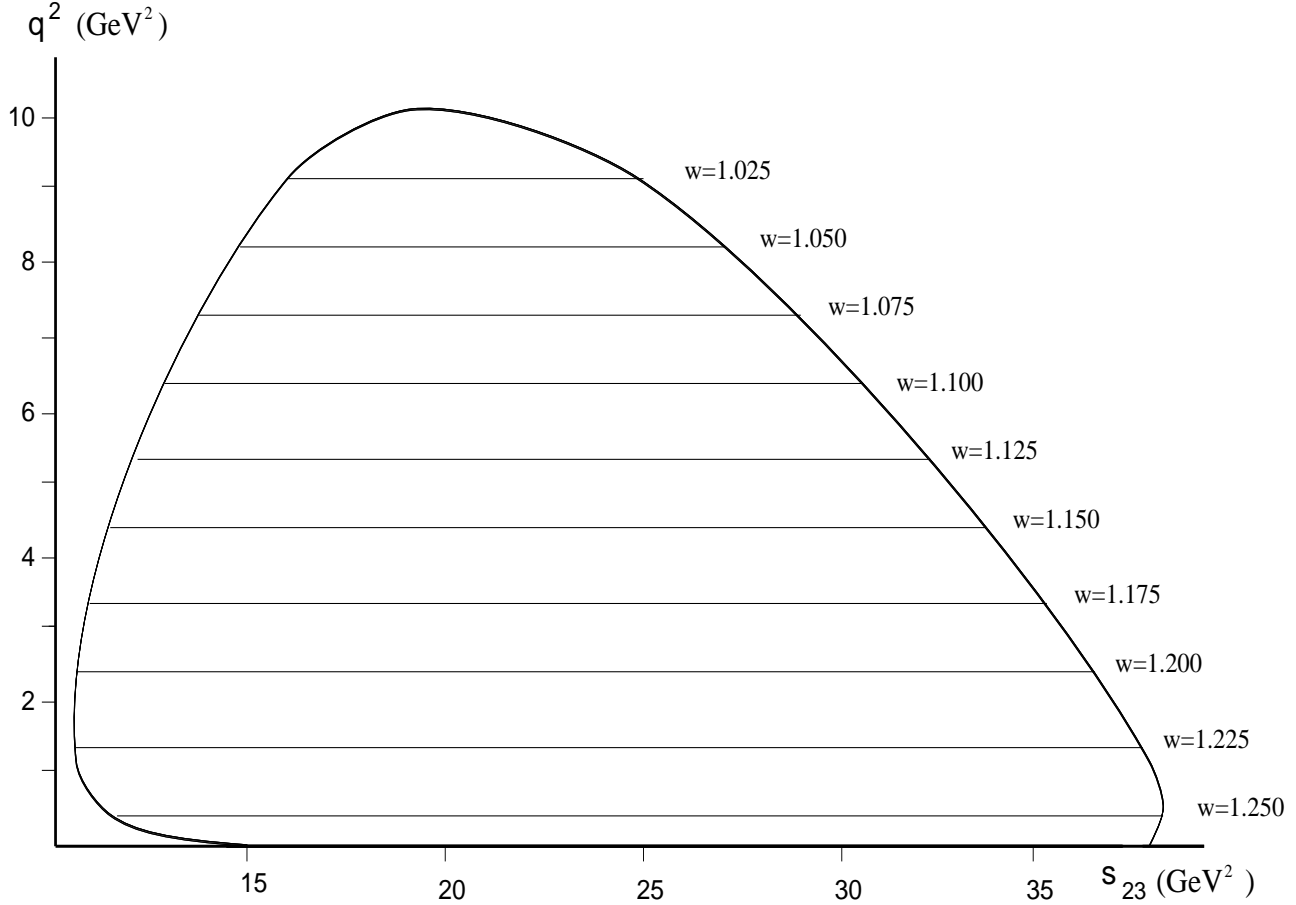,height=12cm,width=17cm}
\caption[\bf Figure 2]{Dalitz plot corresponding to the decay
$B_c{\rightarrow}J/\psi \mu^+\nu$. The leptonic momentum squared is given by
$q^2=(p_{\mu}+p_{\nu})^2=(p_{B_c}-p_{J/\psi})^2$, and
$s_{23}=(p_{J/\psi}+p_{\mu})^2$. The horizontal lines correspond to constant
$w = v_{B_c} \cdot v_{J/\psi}$. The non-recoil point of the $J/\psi$ in the
decaying rest frame has coordinates
$s_{23}\ \simeq \ m_{B_c} m_{J/\psi}\ \simeq\ 19.5$ GeV$^2$ and
$q^2=(m_{B_c}-m_{J/\psi})^2\ \simeq\ 10.1$ GeV$^2$. Notice the linearity
between the velocity transfer $w$ and $q^2$ as a kinematic variable
characterizing the form factors variation for the hadronic transition.}
\end{figure}

\newpage

\begin{figure}[end]
\psfig{figure=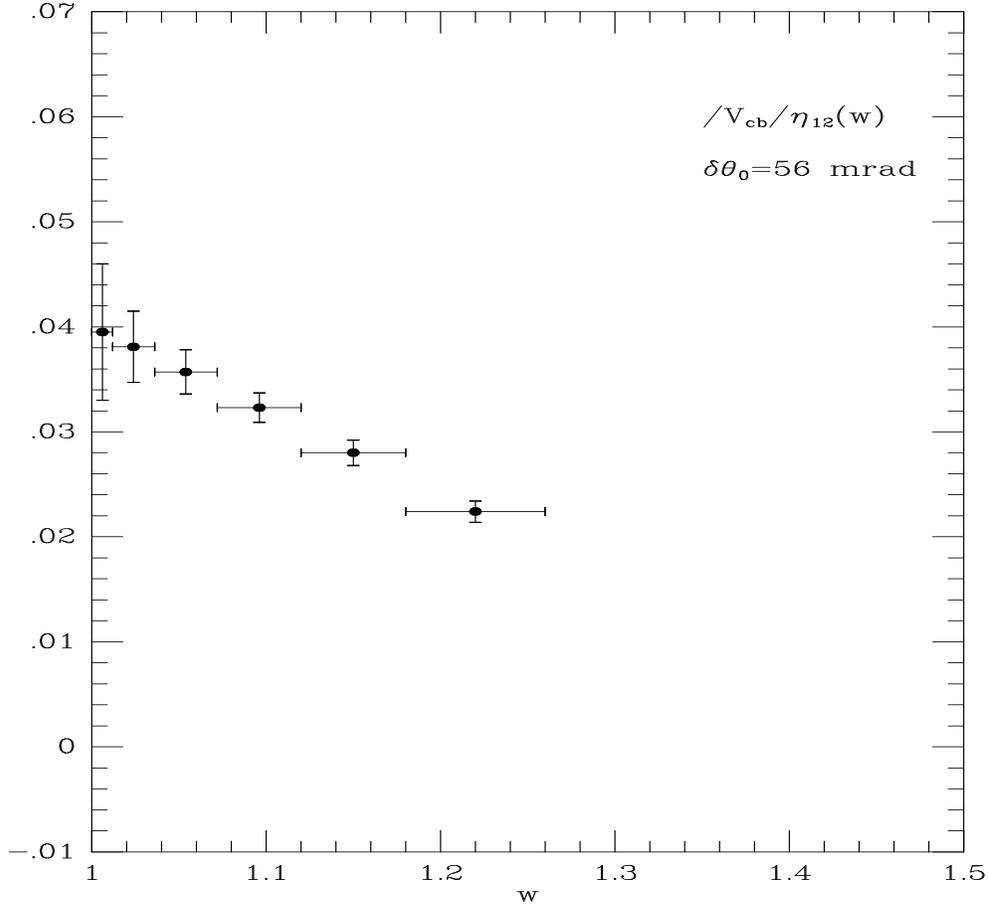,height=15cm,width=15cm}
\caption[\bf Figure 3]{Hypothetical ${\eta}_{12}(w){\cdot}{\mid}V_{cb}{\mid}$
distribution of the decay $B_c{\rightarrow}J/\psi \mu^+\nu$ for a
$500$ events sample tentatively assigned to an integrated luminosity
of $20\ fb^{-1}$ (i.e. two years of running at low luminosity). The
available $w$ range extends over the interval $(1,1.26)$. It was assumed
an spatial resolution for flight-path reconstruction of
${\sigma}\ {\approx}\ 100\ \mu$m and a cut of $n=7$ lifetime units
(${\delta}{\theta}_0\ {\simeq}\ 56$ mrad). Horizontal bars show systematic
errors basically limiting the number of experimental points. Vertical
bars take into account both the
statistical fluctuations of the sampled events in the bin (distributed
according to the expected differential rate for a pseudoscalar to vector
semileptonic decay \cite{neu2}), and the uncertainty
coming from contamination of higher resonances in the $B_c$ decay. Points
line up with ${\eta}'_{12}$ set equal to $-2$.}
\end{figure}

\newpage

\begin{figure}[end]
\centerline{\hbox{
\psfig{figure=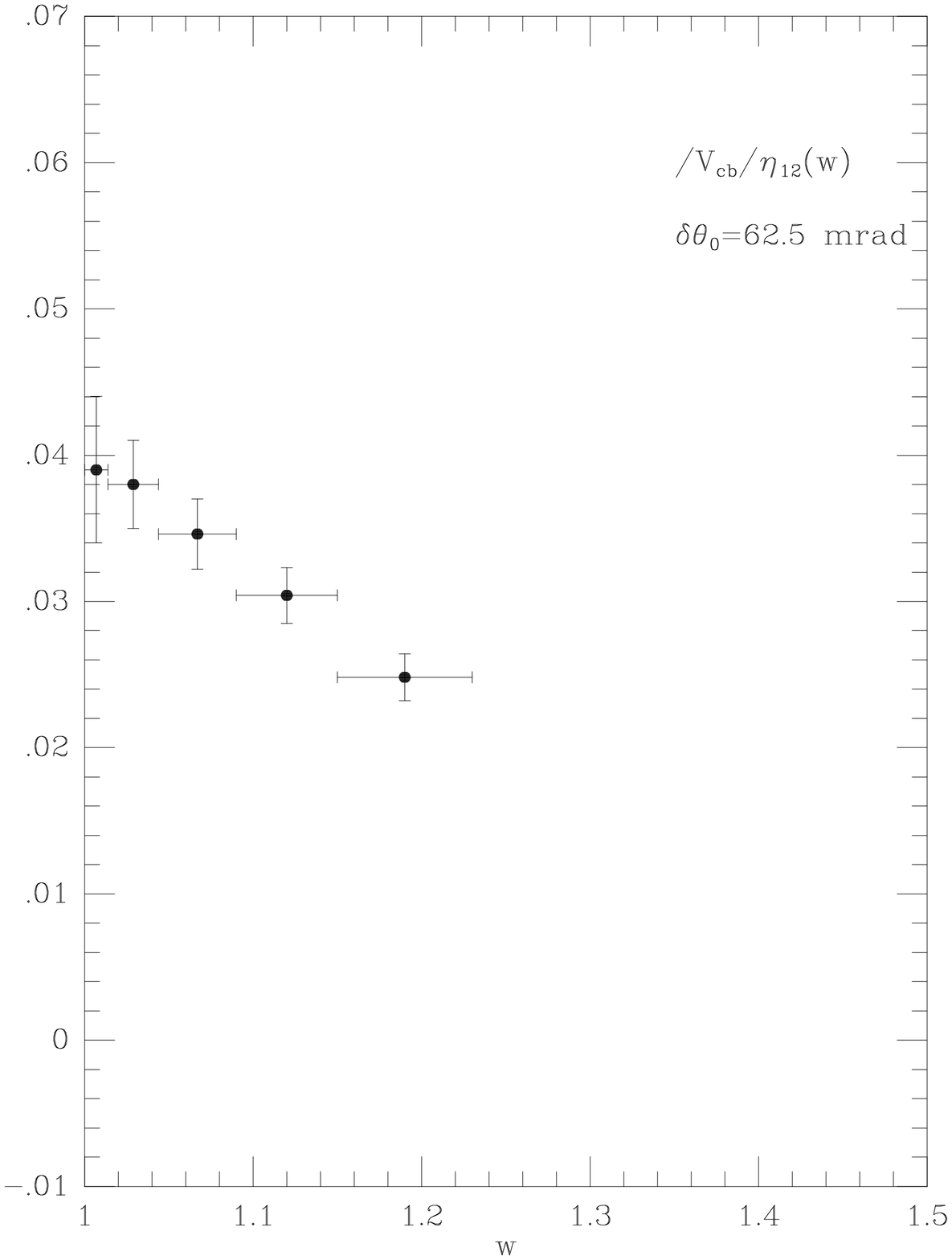,height=10cm,width=10cm}
\psfig{figure=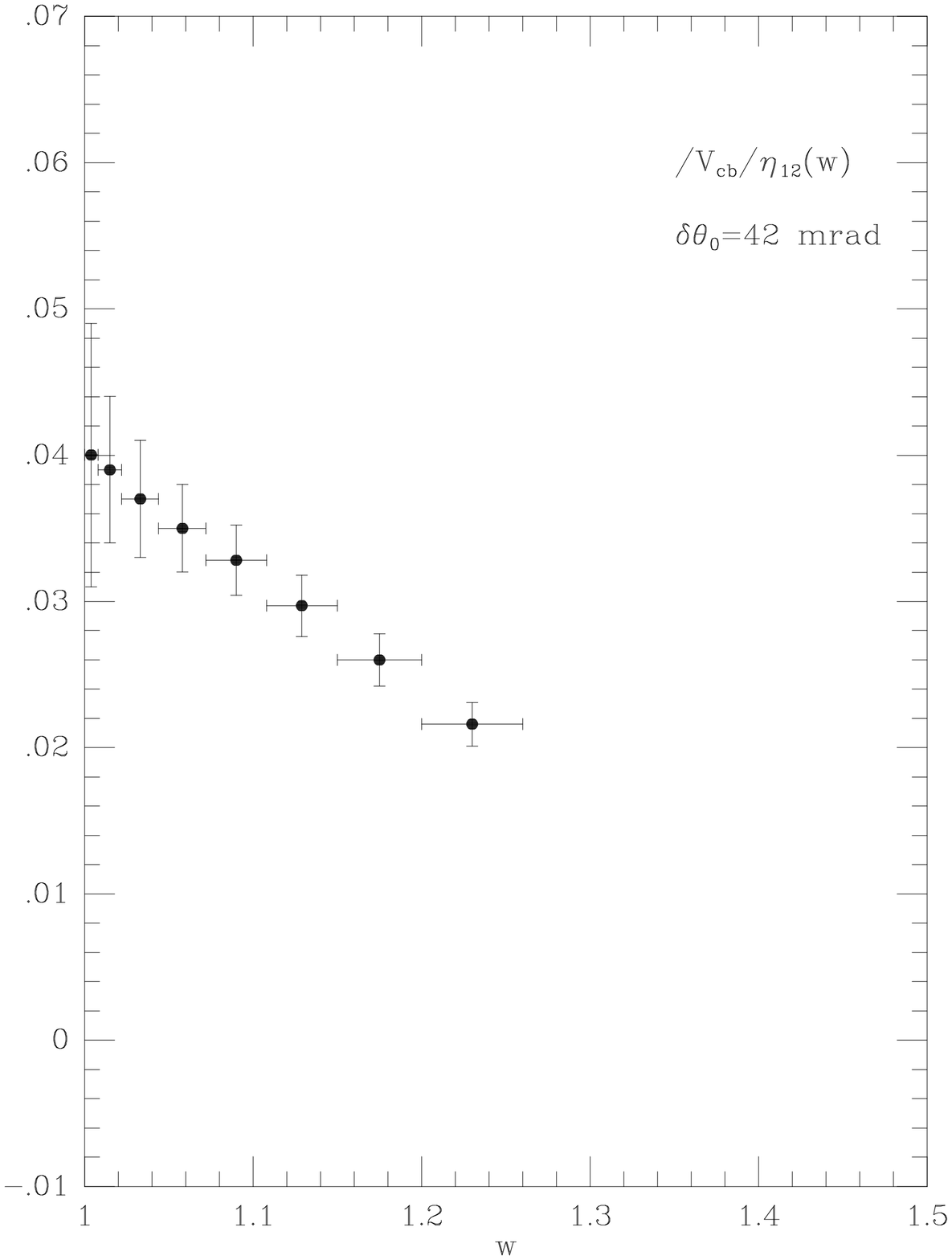,height=10cm,width=10cm}
}}
\end{figure}
\begin{figure}[end]
\centerline{\hbox{
\psfig{figure=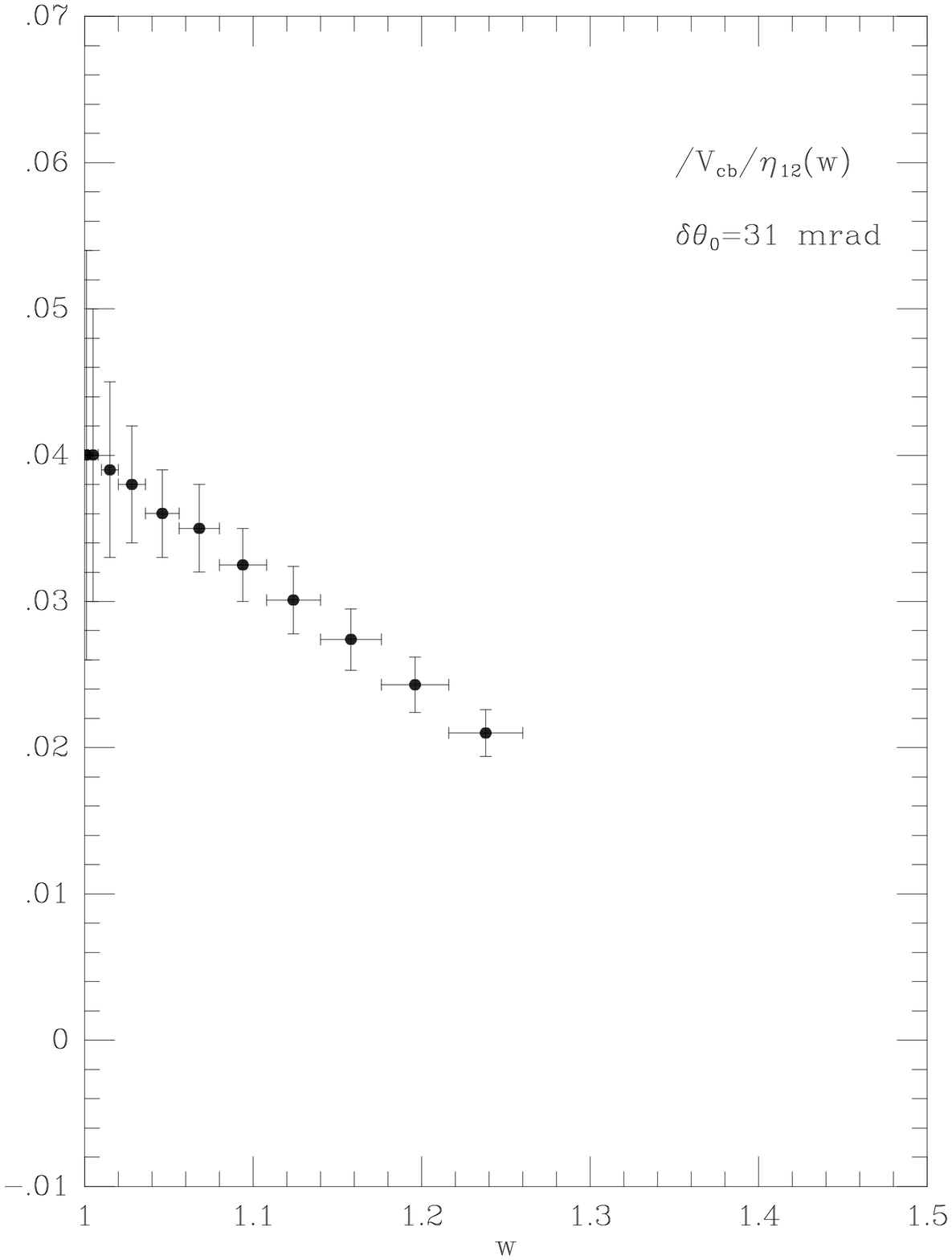,height=10cm,width=10cm}
\psfig{figure=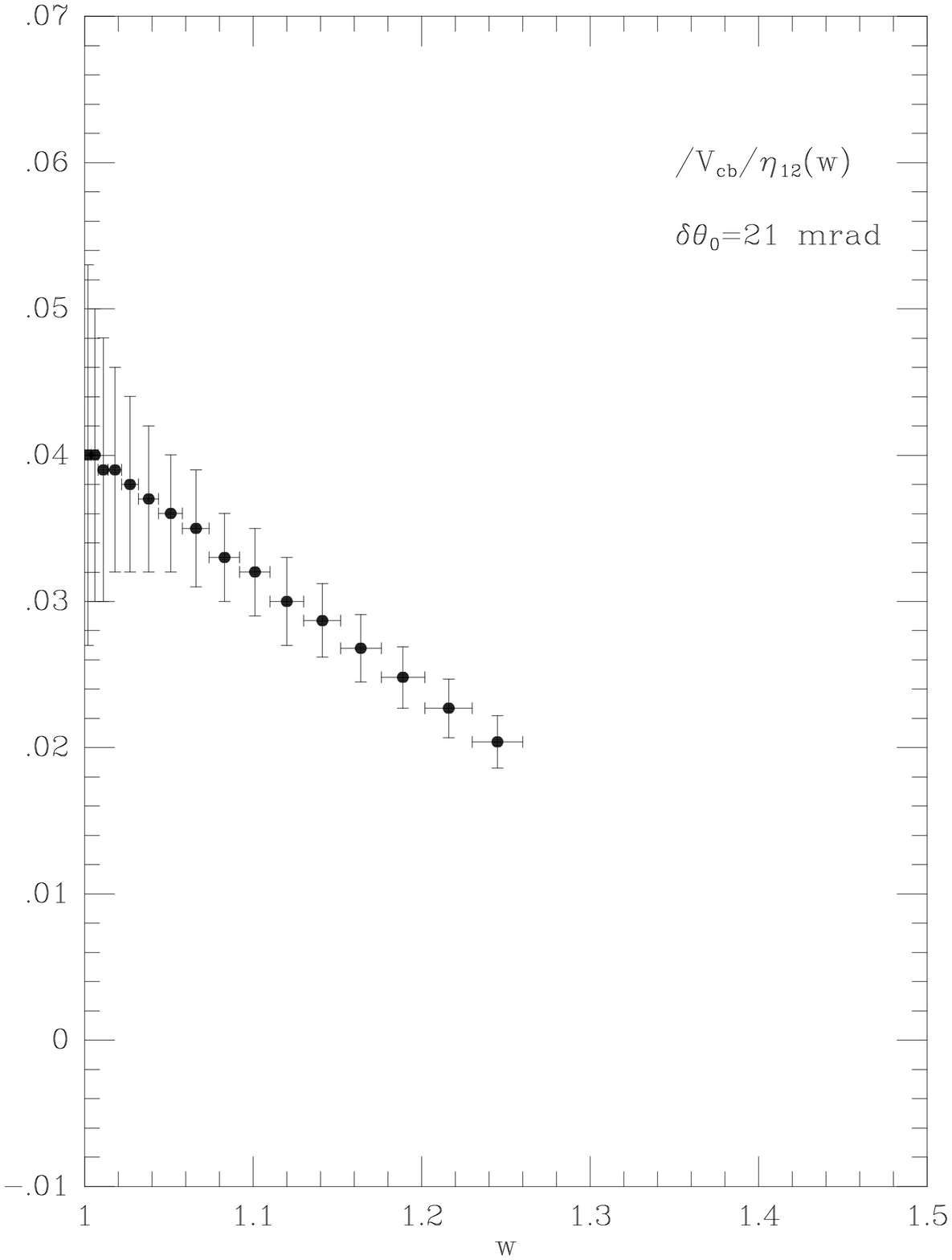,height=10cm,width=10cm}
}}
\caption[\bf Figure 4]{The same as in figure 3 but with different values for
${\delta}{\theta}_0$ corresponding to variations of ${\sigma}$ between
$50$ and $100$ $\mu$m and $\tau$ between $0.5$ and $1$ ps. The improvement
on the number of points near zero recoil can be considerable.}
\end{figure}

\end{document}